# Relativistic dynamics without collisions and conservation laws

Bernhard Rothenstein[1)] and Stefan Popescu[2)]

1) Politehnica University of Timisoara, Physics Department, Timisoara, Romania
2) Siemens AG, Erlangen, Germany

***Abstract***. *We show that the relativistic expressions for momentum and energy as well as the way in which they transform could be derived without involving collisions and conservation laws. Our approach involves relativistic kinematics via the addition law of relativistic velocities.*

In a recent paper Adkins[1] presents an approach to relativistic dynamics which involves the legendary Lewis-Tolman symmetric collision[2] in order to derive the special relativistic expressions for momentum and energy. The author starts with the following assumptions:
- Both the momentum and energy of an object are proportional to the object's mass
- The momentum is proportional to the object's velocity.

Based on these assumptions the author continues with guessed expressions for the relativistic momentum **p** and for the relativistic energy $E$ of an object that moves with velocity **u** of the form:

$$\mathbf{p} = m\mathbf{u} f(u) \qquad (1)$$

$$E = mc^2 g(u) \qquad (2)$$

where $f(u)$ and $g(u)$ are dimensionless functions of the object's speed. These functions are continuous and smooth for $0<u<c$. Beside the fact that the Lewis-Tolman approach is laborious, concentrating more on algebra than on the physics behind it, the assumptions made above are not in an easy grasp for the beginner.

The purpose of our paper is to show that the expressions for relativistic momentum, energy and kinetic energy as well as for their transformation when considered from two inertial reference frames S and S' could be derived without involving collisions and conservation laws. A fundamental role in our derivation is played by the expression for the relativistic velocity transformation written as:

$$u_x = u\cos\theta = \frac{u'_x + V}{1 + \frac{Vu'_x}{c^2}} = \frac{u'\cos\theta' + V}{1 + \frac{Vu'\cos\theta'}{c^2}} \qquad (3)$$

$$u_y = u\sin\theta = \frac{u'_y\sqrt{1-\frac{V^2}{c^2}}}{1+\frac{Vu'_x}{c^2}} = u'\frac{\sqrt{1-\frac{V^2}{c^2}}\sin\theta'}{1+\frac{Vu'}{c^2}\cos\theta'} \qquad (4)$$

where $V$ represents the velocity of S' relative to S pointing in the positive direction of the overlapped axes OX(O'X). Above $u_x = u\cos\theta; u_y = u\sin\theta; u'_x = u'\cos\theta'; u'_y = u'\sin\theta'$ represent the components of object velocities having the magnitudes $u$ and respectively $u'$, with $\theta$ and $\theta'$ representing the angles made by **u** and **u'** with the positive direction of the OX(O'X') axes. The angles $\theta$ and $\theta'$ transform as:

$$\tan\theta = \frac{u_y}{u_x} = \frac{\sqrt{1-\frac{V^2}{c^2}}\sin\theta'}{\cos\theta' + \frac{V}{u'}}. \qquad (5)$$

We underline that (3), (4) and implicitly (5) could be derived without involving the Lorentz-Einstein transformations for the space-time coordinates of the same event.[3]

## 2. Relativistic energy, relativistic momentum, kinetic energy and their transformation without collisions and conservation laws

In our approach we will make use of the fact that the transformation equations (3) and (4) lead to the following relativistic identities:

$$\frac{1}{\sqrt{1-\frac{u^2}{c^2}}} = \frac{1+\frac{Vu'\cos\theta'}{c^2}}{\sqrt{1-\frac{u'^2}{c^2}}\sqrt{1-\frac{V^2}{c^2}}} \qquad (6)$$

$$\frac{u\cos\theta}{\sqrt{1-\frac{u^2}{c^2}}} = \frac{u'\cos\theta' + V}{\sqrt{1-\frac{u'^2}{c^2}}\sqrt{1-\frac{V^2}{c^2}}} \qquad (7)$$

$$\frac{u\sin\theta}{\sqrt{1-\frac{u^2}{c^2}}} = \frac{u'\sin\theta'}{\sqrt{1-\frac{u'^2}{c^2}}}. \qquad (8)$$

Furthermore (6), (7) and (8) remain identities if we multiply both their sides with proper physical quantities like the mass (rest mass) $m$ or the speed of light in empty space $c$.

Multiplying both sides of (6) with $mc^2$ and both sides of (7) and (8) with $m$ and reformatting the result in a suitable way we obtain that:

$$[\frac{mc^2}{\sqrt{1-\frac{u^2}{c^2}}}] = \frac{[\frac{mc^2}{\sqrt{1-\frac{u'^2}{c^2}}}] + V[\frac{mu'_x}{\sqrt{1-\frac{u'^2}{c^2}}}]}{\sqrt{1-\frac{V^2}{c^2}}} \qquad (9)$$

$$[\frac{mu_x}{\sqrt{1-\frac{u^2}{c^2}}}] = \frac{[\frac{mu'_x}{\sqrt{1-\frac{u'^2}{c^2}}}] + V[\frac{m}{\sqrt{1-\frac{u'^2}{c^2}}}]}{\sqrt{1-\frac{V^2}{c^2}}} \qquad (10)$$

and

$$[\frac{mu_y}{\sqrt{1-\frac{u^2}{c^2}}}] = [\frac{mu'_y}{\sqrt{1-\frac{u'^2}{c^2}}}]. \qquad (11)$$

Equations (9), (10) and (11) are genuine transformation equations per se. We make use of the names that physicists (as well trained godfathers) found out for the physical quantities in the brackets allowing for

$$p_x = \frac{mu_x}{\sqrt{1-\frac{u^2}{c^2}}} \qquad (12)$$

and

$$p'_x = \frac{m'u'_x}{\sqrt{1-\frac{u'^2}{c^2}}} \qquad (13)$$

to represent the OX(O'X') components of the momentum detected from S and S', respectively with

$$p_y = \frac{mu_y}{\sqrt{1-\frac{u^2}{c^2}}} \qquad (14)$$

and

$$p'_y = \frac{mu'_y}{\sqrt{1-\frac{u^2}{c^2}}} \qquad (15)$$

representing its OY(O'Y') components. Equation (11) is telling us that the normal components of the momentum have the same magnitude in all inertial reference frames in relative motion. According to the same criteria the relativistic energies detected from S and S' are

$$E = \frac{mc^2}{\sqrt{1-\frac{u^2}{c^2}}} \quad \text{and respectively}$$

$$E' = \frac{mc^2}{\sqrt{1-\frac{u'^2}{c^2}}} \quad (16)$$

With the new notations (9), (10) and (11) become the transformation equations for the components of momentum

$$p_x = \frac{p'_x + \frac{V}{c^2}E'}{\sqrt{1-\frac{V^2}{c^2}}} \quad (17)$$

$$p_y = p'_y \quad (18)$$

and for the energy

$$E = \frac{E' + Vp'_x}{\sqrt{1-\frac{V^2}{c^2}}}. \quad (19)$$

If the object is at rest in S' ($u'=0$) then the observers of this frame measure its rest energy $E_0$ with equation (19) telling us that

$$E = \frac{E_0}{\sqrt{1-\frac{V^2}{c^2}}} \quad (20)$$

in accordance with the fact that the object appears now to move with velocity $V$ relative to S.

Because the single supplementary energy the free object considered so far could posses is its kinetic energy $K$, we obtain that (when detected from S) it will be given by

$$K = E - E_0 = E_0\left[\left(1-\frac{V^2}{c^2}\right)^{-1/2} - 1\right]. \quad (21)$$

A transformation equation for the kinetic energy is proposed in [4].

## 3. Conclusions

Adkins[1] argues that the merit of his approach consists in the fact that it leads to relativistic expressions for both momentum and energy. We believe that our approach is simpler and transparent and furthermore it leads to the transformation equations for momentum and energy advocating for the kinematical roots of relativistic dynamics. Moreover our approach is time saving, an important aspect to be considered when the time allocated to teach physics shrinks dramatically.

**References**
[1]Gregory S. Atkins, "Energy and momentum in special relativity," Am.J.Physics **76**, 1045-1047 (2008) and references therein.
[2]G.N. Lewis and R.C. Tolman, "The principle of relativity and non-Newtonian mechanics," Philos. Mag. **18**, 510-523 (1909)
[3]W.N. Mathews Jr, "Relativistic velocity and acceleration transformations from thought experiments," Am.J.Phys. **73,** 45-51 (2005)
[4]Bernhard Rothenstein and Stefan Popescu, "Transformation equations for the kinetic energy of tardyon and photon via Bertozzi's experiment," Journal Physics Students, **2**, L1 (2008)